# Switchable moiré potentials in ferroelectric WTe$_2$/WSe$_2$ superlattices

Kaifei Kang[1], Wenjin Zhao[2], Yihang Zeng[3], Kenji Watanabe[4], Takashi Taniguchi[4], Jie Shan[1-3], and Kin Fai Mak[1-3]

[1]School of Applied and Engineering Physics, Cornell University, Ithaca, NY, USA
[2]Kavli Institute at Cornell for Nanoscale Science, Ithaca, NY, USA
[3]Laboratory of Atomic and Solid State Physics, Ithaca, NY, USA
[4]National Institute for Materials Science, 1-1 Namiki, 305-0044 Tsukuba, Japan

Email: jie.shan@cornell.edu; kinfai.mak@cornell.edu

**Moiré materials, with superlattice periodicity many times the atomic length scale, have enabled the studies of strong electronic correlations and band topology with unprecedented tunability [1-5]. However, nonvolatile control of the moiré potentials, which could allow on-demand switching of the superlattice effects, has not been achieved to date. Here we demonstrate the switching of the correlated and moiré band insulating states and the associated nonlinear anomalous Hall effect by the ferroelectric effect. This is achieved in a ferroelectric WTe$_2$ bilayer of the T$_d$ structure with a centered-rectangular moiré superlattice induced by interfacing with a WSe$_2$ monolayer of the H structure. The results can be understood in terms of polarization-dependent charge transfer between two WTe$_2$ monolayers, which possess very different moiré potential depths; ferroelectric switching thus turns on/off the superlattice. Our study demonstrates the potential of creating new functional moiré materials by incorporating intrinsic symmetry-breaking orders [6-8].**

**Main**

Two-dimensional ferroelectric materials have attracted much recent interest due to their potentials in non-volatile memory device applications [9]. In addition to intrinsic ferroelectric materials [10-12], such as CuInP$_2$S$_6$ and few-layer T$_d$-WTe$_2$, ferroelectricity has also been realized recently in moiré heterostructures [13-17], enabling non-volatile control of correlated electronic states and superconductivity. However, non-volatile electrical switching of the moiré trapping potentials for electrons and the associated superlattice effects has not been realized to date. Here we achieve this goal by utilizing the intrinsic ferroelectricity in angle-aligned bilayer T$_d$-WTe$_2$/monolayer H-WSe$_2$ moiré heterostructures.

Monolayer WTe$_2$ of the T$_d$ crystal structure consists of a layer of W atoms between two layers of Te atoms in the distorted octahedral coordination [18]. The structure is centrosymmetric with a mirror line along the crystal b-axis. At charge neutrality, the material is a quantum spin Hall insulator [19-21]. Bilayer T$_d$-WTe$_2$ is made of two antiparallel monolayers; the structure becomes non-centrosymmetric with a polar axis [22] (inclined from the out-of-plane direction). The presence of the polar axis stabilizes ferroelectricity [12] and generates Berry curvature dipoles in the momentum space [23] responsible for the nonlinear anomalous Hall effect [24-26]. Bilayer T$_d$-WTe$_2$ is a topologically trivial insulator. Figure 1c illustrates its schematic band structure with a flat valence band centered at the Γ-point of the Brillouin zone and dispersive conduction bands away from the Γ point [27,28]. The bands are layer-polarized by the spontaneous out-of-plane electric polarization.

**WTe$_2$/WSe$_2$ moiré superlattice**
We introduce moiré potentials to bilayer WTe$_2$ by adding a WSe$_2$ monolayer on it (Fig. 1a). The WSe$_2$ monolayer is of the H crystal structure, in which the W atoms are between the Se atoms in the trigonal prismatic coordination. The zigzag or armchair crystal axis of WSe$_2$ is aligned to the crystal a-axis of WTe$_2$. A nearly centered-rectangular moiré superlattice is expected with periods 6.9 nm and 6.4 nm along the a- and b-axis of WTe$_2$, respectively. These values are estimated from the lattice mismatch between the two materials (5% and 9%) along the two axes. The corresponding moiré density, $n_M$, is about $4.5 \times 10^{12} \text{cm}^{-2}$. These estimates are in good agreement with the experimental moiré pattern of a typical aligned WTe$_2$/WSe$_2$ heterostructure obtained by piezoresponse force microscopy (Fig. 1b, see Methods and Extended Data Fig. 1 for details). The two lattice vectors form an angle of about 84 degrees. The moiré lattice is robust and insensitive to twist-angle variations in the sample as a result of the large lattice mismatch. The nearly centered-rectangular lattice also contrasts with the triangular and honeycomb moiré lattices examined by most of the existing studies to date [1-5].

The heterostructure has a type-I band alignment with both the conduction band minimum and valence band maximum from bilayer WTe$_2$ (Ref. [29]). The WSe$_2$ layer remains charge-neutral at all times; it only provides a moiré potential responsible for the flat bands in the WTe$_2$ layer. Moreover, the top WTe$_2$ layer directly interfacing with WSe$_2$ is expected to experience a substantially stronger moiré potential compared to the bottom WTe$_2$ layer; the moiré bands for the top layer are expected to be much flatter (Fig. 1c). Such layer-sensitive moiré effects in bilayer WTe$_2$, combined with its intrinsic ferroelectric order, form the basis for ferroelectric control of the moiré potential in this study.

Below we demonstrate the effects of ferroelectric switching on the correlated insulating states and the Berry curvature dipole distribution in the correlated flat bands. The heterostructure is encapsulated between top and bottom boron nitride and graphite gates for independent controls of the charge carrier density $\nu$ (in units of $n_M$) and the electric field $E$ perpendicular to the sample plane (positive field defined as pointing from WTe$_2$ to WSe$_2$). See Methods for details on device fabrications, calibration of moiré density and measurements.

To contrast the behavior of bilayer WTe$_2$ devices, we first examine the density and electric field-dependent transport properties in an angle-aligned monolayer WTe$_2$/monolayer WSe$_2$ heterostructure (Fig. 1e). Since monolayer WTe$_2$ is a quantum spin Hall insulator [19-21], we fabricate a Corbino geometry device in order to access the bulk resistivity of the material (Methods and Extended Data Fig. 2). Because the electrical contact resistance of this device increases substantially below 30 K, we limit our measurements down to 30 K. In Fig. 1e, we observe resistance peaks at filling factors of $\nu = 0, -1, -2$. They correspond to the band insulating state when WTe$_2$ is at charge neutrality ($\nu = 0$), the moiré band insulating state when the first moiré valence band is fully filled with holes ($\nu = -2$), and the correlated insulating state (likely a Mott insulator) for a half-filled moiré band ($\nu = -1$). The insulating states at $\nu = -1, -2$ confirms the formation of flat moiré valence bands in monolayer WTe$_2$. In contrast, no insulating state is observed at $\nu = 1, 2$ on the electron doping side (see also Fig. 2, 3). The result demonstrates the negligible moiré effects on the conduction bands possibly due to the small conduction band mass in WTe$_2$ (Ref. [27,28]) and the small moiré periods in the WTe$_2$/WSe$_2$ heterostructure. Also, all of the observed insulating states show weak and non-hysteretic electric field dependence, consistent with the absence of ferroelectricity in monolayer WTe$_2$ (Ref. [12]).

**Switchable moiré potentials by ferroelectricity**
Next we examine the density and electric field dependent resistance in a bilayer WTe$_2$ device under forward (from negative to positive) and backward (from positive to negative) scans of the electric field in Fig. 2a and 2b, respectively. The Corbino geometry is no longer needed because bilayer WTe$_2$ is a topologically trivial insulator [27,28]. The electrical contacts are also better than the monolayer device; lower temperature measurements are possible. Similar to the monolayer device, we observe insulating states at $\nu = -1, -2$ for hole doping, a charge-neutral insulating state at $\nu = 0$ and no moiré insulating states for electron doping. Unlike the monolayer device, strong electric field dependence is observed. In particular, robust insulating states at $\nu = -1, -2$ are observed mostly for $E \lesssim 0.3$ V/nm in the forward scan (Fig. 2a) and for $E \lesssim 0.1$ V/nm in the backward scan (Fig. 2b). A sharp resistance jump and an electric field hysteresis are also observed near these critical fields, as exemplified by the electric field dependent resistance at $\nu = -1$ in Fig. 2c (similar data at other filling factors are shown in Extended Data Fig. 3). In addition to the hysteresis, a non-monotonic electric field dependence is also observed (the origin is discussed in Methods). The results clearly demonstrate the effects of ferroelectric switching (in bilayer WTe$_2$) on the moiré physics. The center of the ferroelectric switching is shifted from 0 V/nm to ~ 0.2 V/nm and disperses with hole doping because of the built-in electric field in the asymmetric heterostructure (WSe$_2$ is on top of WTe$_2$).

We further illustrate the effects of ferroelectric switching by plotting the density dependent resistance near the center of the hysteresis (at fixed $E = 0.2$ V/nm) for both forward and backward field scans (Fig. 2d). The two cases represent the two remnant states of spontaneous polarization $P$ in the ferroelectric WTe$_2$: positive polarization ($P > 0$, pointing from WTe$_2$ to WSe$_2$) for backward scan and negative polarization ($P < 0$, pointing from WSe$_2$ to WTe$_2$) for forward scan. Insulating states at $\nu = -1, -2$ associated with the moiré valence band are observed only for $P < 0$; only the $\nu = 0$ insulating state remains for $P > 0$. An on-off ratio of ~ 10 for the insulating states at $\nu = -1, -2$ is achieved upon ferroelectric switching (inset).

The results above can be understood based on the interlayer transfer of mobile holes in bilayer WTe$_2$ driven by its spontaneous ferroelectric polarization, which originates from the electric-field-switchable ionic displacements in the material [30]. When positive polarization ($P > 0$) is prepared, the internal electric field from the displaced ions (or bound charges) within the bilayer WTe$_2$ is negative; the field splits the electronic bands of the top and bottom WTe$_2$ layer via the Stark effect and pushes the mobile holes to the bottom layer (Fig. 1d). The reverse happens when negative polarization ($P < 0$) is prepared (Fig. 1d). Because holes in the top layer interfacing directly with the WSe$_2$ layer experience a strong moiré potential, robust insulating states at $\nu = -1, -2$ are observed for $P < 0$. The absence of these insulating states for $P > 0$ shows the negligible moiré potential in the bottom WTe$_2$ layer.

Note that the total electric polarization is a sum of both the bound charge and mobile hole contributions; the two contributions point in opposite directions. The bound charge contribution dominates over the entire hole doping range; this guarantees nearly complete (i.e. 100 %) interlayer hole transfer upon ferroelectric switching and the emergence of the moiré insulating states at $\nu = -1, -2$ for $P < 0$. Otherwise, the more mobile holes in the bottom layer would electrically short the bilayer conduction irrespective of the sign of $P$; moiré insulating states would not have been observed.

**Variable-range hopping transport**
To gain further insights into how ferroelectric switching influences the charge transport in bilayer WTe$_2$, we perform temperature dependence studies for both $P < 0$ and $P > 0$, which correspond to with and without moiré effects, respectively. Figure 3a and 3b show the density dependent resistance at varying temperatures (5 – 90 K) for the two states. Metallic transport (decreasing resistance with decreasing temperature) is observed only on the electron doping side, where moiré physics is not observable; a metal-insulator transition is also observed near $\nu = 0.7$. For the entire hole doping range, the material exhibits insulator-like behavior (increasing resistance with decreasing temperature) even when there is no moiré physics for $P > 0$. The $\nu = -1, -2$ insulating states for $P < 0$ survive up to about 70 K.

The temperature dependent resistance at $\nu = -1$ for both $P < 0$ and $P > 0$ is shown in Fig. 3c (similar dependence at other densities is shown in Extended Data Fig. 4). For temperatures below about 100 K, the resistance $R$ scales with the temperature $T$ according to the Efros-Shklovskii variable-range hopping model [31,32], $R \propto \exp(T_0/T)^{1/2}$, for over two orders of magnitude in $R$. The characteristic temperature $T_0 = \frac{2.8\, e^2}{4\pi\varepsilon\varepsilon_0 k_B \xi}$ is inversely proportional to the localization length $\xi$ of holes in the bilayer WTe$_2$ (here $e$, $\varepsilon_0$, $\varepsilon$ and $k_B$ denote the electron charge, the vacuum permittivity, the dielectric constant of the substrate and the Boltzmann constant, respectively). The density dependence of $T_0$ is extracted in Fig. 3d. The temperature $T_0$ is substantially enhanced over the entire hole doping range for $P < 0$; additional enhancement at the $\nu = -1, -2$ insulating states is also observed. The results demonstrate the additional localization of holes that shortens $\xi$ ($\approx 35$ nm for $T_0 = 300$ K) when the moiré potential is turned on for $P < 0$.

**Nonlinear anomalous Hall effect**
Finally, we examine the effects of ferroelectric switching on the topological transport properties. The presence of an in-plane polar axis (along the mirror line of WTe$_2$) in the non-centrosymmetric centered-rectangular moiré lattice allows the generation of an out-of-plane magnetization under a bias current perpendicular to the mirror line [23,25]. The current-induced magnetization can in turn induce an anomalous Hall effect probed by the same bias current. The anomalous Hall voltage thus scales with the bias current squared, giving rise to a nonlinear anomalous Hall effect (NAHE) [23-26]. The effect provides a powerful probe of the Berry curvature dipoles (an imbalance in the Berry curvature occupations) in moiré materials [33,34].

To measure the NAHE in the WTe$_2$/WSe$_2$ moiré heterostructure, we bias an ac current ($I$) perpendicular to the mirror line of the moiré lattice and simultaneously measure the first-harmonic longitudinal voltage ($V_\parallel$) and the second-harmonic transverse voltage ($V_\perp^{2\omega}$) (Methods). Figure 4a and 4b show the density and electric field dependence of $V_\perp^{2\omega}/I^2$ under forward and backward scans in the electric field, respectively. The quantity provides a measure of the nonlinear anomalous Hall response (Extended Data Fig. 5). The measurement temperature is kept above 25 K for good electrical contacts. The results are correlated with the behavior for the resistance in Fig. 2b and 2c. In particular, strong NAHE with sign reversal in $V_\perp^{2\omega}/I^2$ is observed at $\nu = -1, -2$ mainly for $E \lesssim 0.3$ V/nm in the forward scan and for $E \lesssim 0.1$ V/nm in the backward scan; or else only the $\nu = 0$ region shows strong NAHE; there is also a sharp jump in $V_\perp^{2\omega}/I^2$ and an electric field hysteresis near these critical electric fields. We extract the density dependent $V_\perp^{2\omega}/I^2$ for the two remnant states of spontaneous polarization $P < 0$ and $P > 0$ in Fig. 4c. (Temperature dependence is shown in Extended Data Fig. 6.) Strong NAHE with sign reversal in $V_\perp^{2\omega}/I^2$ is observed at $\nu = 0, -1, -2$ for $P < 0$; only the characteristic response at $\nu = 0$ remains for $P > 0$.

The results demonstrate the ferroelectric switching of the NAHE. In the absence of the moiré potential ($P > 0$), the Berry curvature and its dipole are concentrated near the WTe$_2$ conduction and valence band edges; the NAHE is strong near charge neutrality ($\nu = 0$); the sign reversal in $V_\perp^{2\omega}/I^2$ reflects the opposite Berry curvature dipoles for the conduction and valence bands [24,27]. When flat bands are formed due to the moiré potential ($P < 0$), the Berry curvature and its dipole are redistributed within the moiré Brillouin zone [35]; the small moiré band gap is expected to greatly enhance the Berry curvature effects [35,36]. In the strong correlation limit (Coulomb repulsion exceeds the electronic bandwidth), Hubbard bands are formed; Berry curvatures of opposite signs are expected to concentrate near the Hubbard band edges, resulting in the characteristic NAHE response observed at $\nu = -1, -2$. Interestingly, the sign reversal in $V_\perp^{2\omega}/I^2$ at $\nu = -1, -2$ resembles the resets of the Hall density at integer fillings in graphene moiré materials [37]. The resets reflect the changing carrier type upon doping an insulating state.

**Conclusions**

In conclusion, we have demonstrated electrical switching of the moiré potential by ferroelectricity in angle-aligned bilayer WTe$_2$/monolayer WSe$_2$; the heterostructure forms a centered-rectangular moiré lattice. The emergence of moiré insulating states at integer filling factors is fully controlled by the spontaneous polarization direction. Ferroelectric switching can also redistribute the Berry curvature dipole in the momentum space, giving rise to a strong and sign-reversed NAHE at integer fillings when the moiré potential is turned on. Although the experimental results can be qualitatively understood based on the formation of Hubbard bands, the distribution of Berry curvature dipole within the Hubbard bands in the strong correlation limit (versus that within the single-particle moiré bands in the non-interacting limit) is an open question for future studies; understanding the NAHE in the hopping transport regime [38] (Fig. 3 and Extended Data Fig. 6) also requires further investigations.

**Methods**
**Device fabrication**
T$_d$-WTe$_2$/H-WSe$_2$ moiré heterostructures were fabricated using the standard layer-by-layer stacking method [39]. In short, air-sensitive monolayer and bilayer WTe$_2$ flakes were exfoliated and identified based on their optical contrast on Si substrates inside a nitrogen-filled glovebox. The a- and b-axis of the WTe$_2$ flakes were determined according to the sharp sample edges. Monolayer WSe$_2$ flakes were exfoliated under ambient conditions. The WSe$_2$ crystal axes were determined by polarization- and angle-resolved second-harmonic generation spectroscopy [40]. The zigzag or armchair axis of WSe$_2$ was aligned to the crystal a-axis of WTe$_2$ during the stacking process. To complete the devices, we first defined platinum electrodes in the Hall-bar geometry on a hexagonal boron nitride (hBN)/graphite heterostructure (that serves as a bottom gate) using electron-beam lithography. We then used a polymer stamp consisting of a polycarbonate (PC) film on polydimethylsiloxane (PDMS) to sequentially pick up flakes of graphite, hBN, a WSe$_2$ monolayer, and a WTe$_2$ monolayer or bilayer. The finished heterostructure on the stamp was released onto the pre-patterned platinum electrodes at 180°C. The polymer stamp was then dissolved in chloroform. To fabricate Corbino geometry devices, we further inserted a thin hBN spacer between the moiré heterostructure and the platinum electrodes so that the WTe$_2$ sample edges do not touch the metal electrodes [41,42]. See Extended Data Fig. 2 for the device geometry.

**Linear and nonlinear electrical measurements**
Electrical transport measurements were carried out in an Oxford Teslatron cryostat down to 1.8 K. Both linear and nonlinear electrical measurements were performed using the standard lock-in technique. Specifically, current with constant amplitude of 0.5 µA and modulation frequency of 17 Hz was biased along the crystal a-axis of WTe$_2$. The longitudinal voltage drop at the same frequency ($V_\parallel$) and the Hall voltage drop at the second-harmonic frequency ($V_\perp^{2\omega}$) were simultaneously recorded by lock-in amplifiers (Stanford Research SR830). In general, the contact resistance increases substantially at low temperatures; we therefore limited the nonlinear anomalous Hall measurements to temperatures above 25 K. The nonlinear anomalous Hall response was verified to be independent of the excitation frequency (Extended Data Fig. 5). In addition, we also measured the longitudinal resistance with current biased along the crystal b-axis of WTe$_2$ (Extended Data Fig. 7). The linear transport behavior is largely independent of the bias current direction.

The carrier density ($n$) and moiré density ($n_M$) of the device shown in the main text were calibrated using the parallel-plate capacitor model: $n = \frac{\varepsilon\varepsilon_0 V_{tg}}{d_{tg}} + \frac{\varepsilon\varepsilon_0 V_{bg}}{d_{bg}}$. Here $d_{tg} \approx 4$ nm and $d_{tg} \approx 22$ nm are the top and bottom hBN gate dielectric thicknesses measured by atomic force microscopy (AFM); $V_{tg}$ and $V_{bg}$ are the top and bottom gate voltages; $\varepsilon_0$ is the vacuum permittivity; and $\varepsilon \approx 3$ is the hBN dielectric constant. A hole density of $(5.0 - 5.5) \times 10^{12}$ cm$^{-2}$ at $\nu = -1$ was obtained using the applied gate voltages relative to those at $\nu = 0$; the value is about 10-20 % larger than the moiré density ($4.5 \times 10^{12}$ cm$^{-2}$) estimated using the known lattice parameters (see main text). The discrepancy may arise from misalignment of the WTe$_2$ and WSe$_2$ crystal axes (up to about 2 degrees) as well as identification of the gate voltage corresponding to zero doping density.

**Piezoresponse force microscopy (PFM)**
To characterize the moiré pattern using PFM, we fabricated an angle-aligned monolayer T$_d$-WTe$_2$/H-WSe$_2$ heterostructure on a polymer stamp consisting of PC on PDMS. Flakes of hBN, monolayer WTe$_2$ and monolayer WSe$_2$ were sequentially picked up by the polymer stamp. We then flipped the stamp for the PFM characterization with WSe$_2$ on the top surface to protect WTe$_2$ from oxidation. PFM was performed using a Bruker Veeco Icon AFM with Oxford Instrument Asylum Research ASYELEC-01 Ti/Ir coated silicon probes. An ac bias voltage with magnitude of 4 V and frequency of 680 kHz was applied to drive the probe during the scan. To improve the signal-to-noise ratio of the obtained moiré pattern, the raw PFM images were Fourier filtered with a threshold of 72% of the maximum intensity (Extended Data Fig. 1).

**Non-monotonic electric-field dependent resistance**

Extended Data Fig. 3 shows the electric-field dependent longitudinal resistance for both field scan directions at selected filling factors. We focus our discussions on $\nu = 0$ and 1, at which high out-of-plane electric fields can be applied. Non-monotonic electric field dependence of the resistance is observed. In particular, resistance peaks are observed at electric fields $E \approx -0.3$ V/nm and 0.55 V/nm for $\nu = 0$ and at $E \approx 0.1$ V/nm and 0.65 V/nm for $\nu = 1$.

To understand the observation, we consider the band structure of and the interlayer charge distribution in bilayer WTe$_2$ under an applied electric field $E$. In the presence of a spontaneous polarization $P$ (dominated by the bound charge contribution), the total electric field is given by $E_{tot} = E - P/\chi\varepsilon_0$ ($\chi$ and $\varepsilon_0$ denote the electric susceptibility and the vacuum permittivity, respectively). At $\nu = 0$ (charge neutrality), the resistance reflects the charge gap size of bilayer WTe$_2$. The gap size is the largest for $E_{tot} = 0$ and decreases linearly with $E_{tot}$ due to the interlayer Stark effect [27] (Fig. 1c). The maximum gap size is therefore expected at $E = \pm |P|/\chi\varepsilon_0$, which is consistent with the observed resistance peaks at $E \approx -0.3$ V/nm and 0.55 V/nm in Extended Data Fig. 3. The asymmetry in the observed electric fields is caused by the heterostructure asymmetry, which shifts the center of the electric-field dependence by about 0.2 V/nm. Away from these two fields, the gap size decreases with $E_{tot}$, and the resistance decreases.

On the other hand, the resistance at $\nu = 1$ is largely determined by the location of the doped holes, specifically, high (low) resistance for holes residing in the top (bottom) WTe$_2$ layer. For $P < 0$, a negative electric field ($E < 0$) pushes the doped holes away from the WSe$_2$/WTe$_2$ interface, thus reducing the moiré effects (Fig. 1d); the resistance decreases monotonically (Extended Data Fig. 3). For $P > 0$, an electric field $E > 0.4$ V/nm pushes the doped holes towards the WSe$_2$/WTe$_2$ interface, therefore enhancing the moiré effects (Fig. 1d); the resistance increases and peaks at $E \approx 0.65$ V/nm, beyond which the charge gap decreases due to the interlayer Stark effect (as discussed above). Finally, hysteretic switching behavior is observed for $0 < E < 0.4$ V/nm; the resistance is dependent on the direction of $P$. The center of the ferroelectric switching is shifted by about 0.2 V/nm because of the heterostructure asymmetry.

**Acknowledgements**
We thank Angel Rubio, Yang Zhang and Jin Zhang for fruitful discussions.

**Author contributions**
K.K. fabricated the devices. K.K. and W.Z. performed the electrical measurements. Y.Z. and K.K. performed the PFM measurements. K.K. analyzed the data. K.W. and T.T. grew the hBN crystals. K.K., J.S. and K.F.M. designed the scientific objectives, oversaw the project, and co-wrote the manuscript. All authors discussed the results and commented on the manuscript.

# Figures

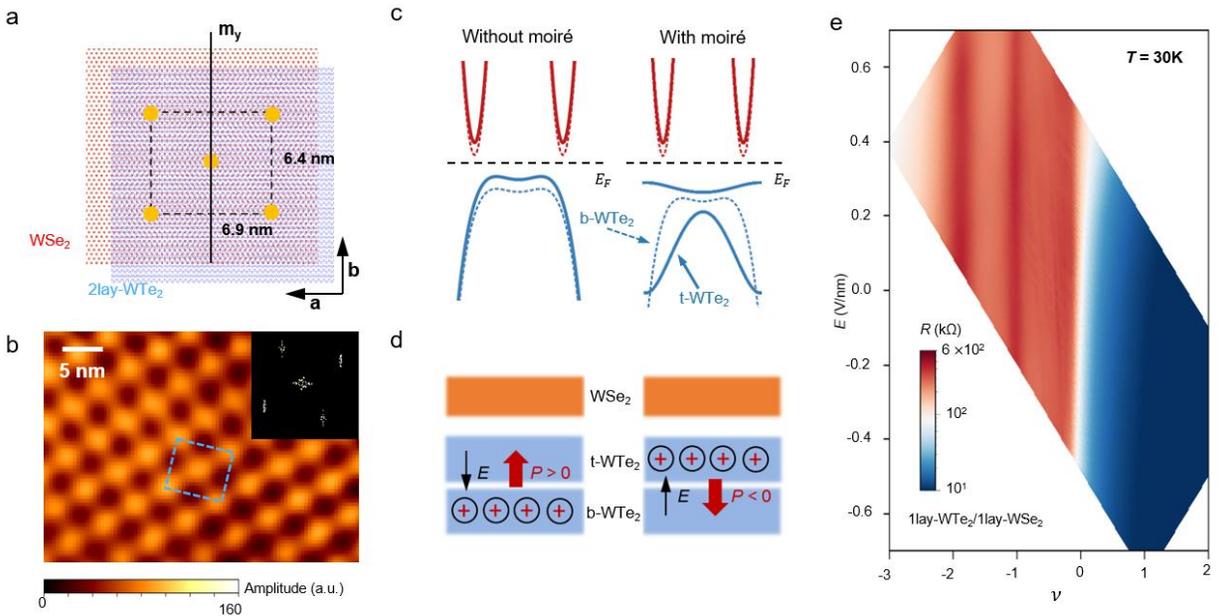

**Figure 1 | Bilayer T$_d$-WTe$_2$/monolayer H-WSe$_2$ moiré heterostructure. a,** Moiré superlattice for angle-aligned bilayer T$_d$-WTe$_2$/monolayer H-WSe$_2$ heterostructure. Only the W atoms are shown. Yellow dots denote the high-symmetry MM (W on W) sites, which form a centered-rectangular lattice with lattice constants of 6.9 nm along the crystal a-axis of WTe$_2$ and 6.4 nm along the crystal b-axis of WTe$_2$ (the mirror line m$_y$). The dashed rectangle marks the moiré unit cell. The black arrows mark the crystal a and b axes of bilayer WTe$_2$. **b,** PFM image of a typical moiré heterostructure. The dashed rectangle marks the moiré unit cell. Inset shows the Fourier transform of the PFM image. **c,** Schematics for the bilayer WTe$_2$ band structure with (right) and without (left) moiré effects. Red and blue colors correspond to conduction and valence bands, respectively. Right: moiré bands are formed on the top WTe$_2$ (t-WTe$_2$) layer only (solid blue curves). The bottom WTe$_2$ (b-WTe$_2$) layer experiences negligible moiré potential; no flat bands are formed (dashed lines). The horizontal black dashed line denotes the Fermi level. **d,** Polarization-dependent hole (red cross) distributions in bilayer WTe$_2$. P and E are, respectively, the spontaneous polarization and its corresponding internal electric field. **e,** Electric field and filling factor dependence for the longitudinal resistance of angle-aligned monolayer T$_d$-WTe$_2$/monolayer H-WSe$_2$ heterostructure. The measurement temperature is 30 K.

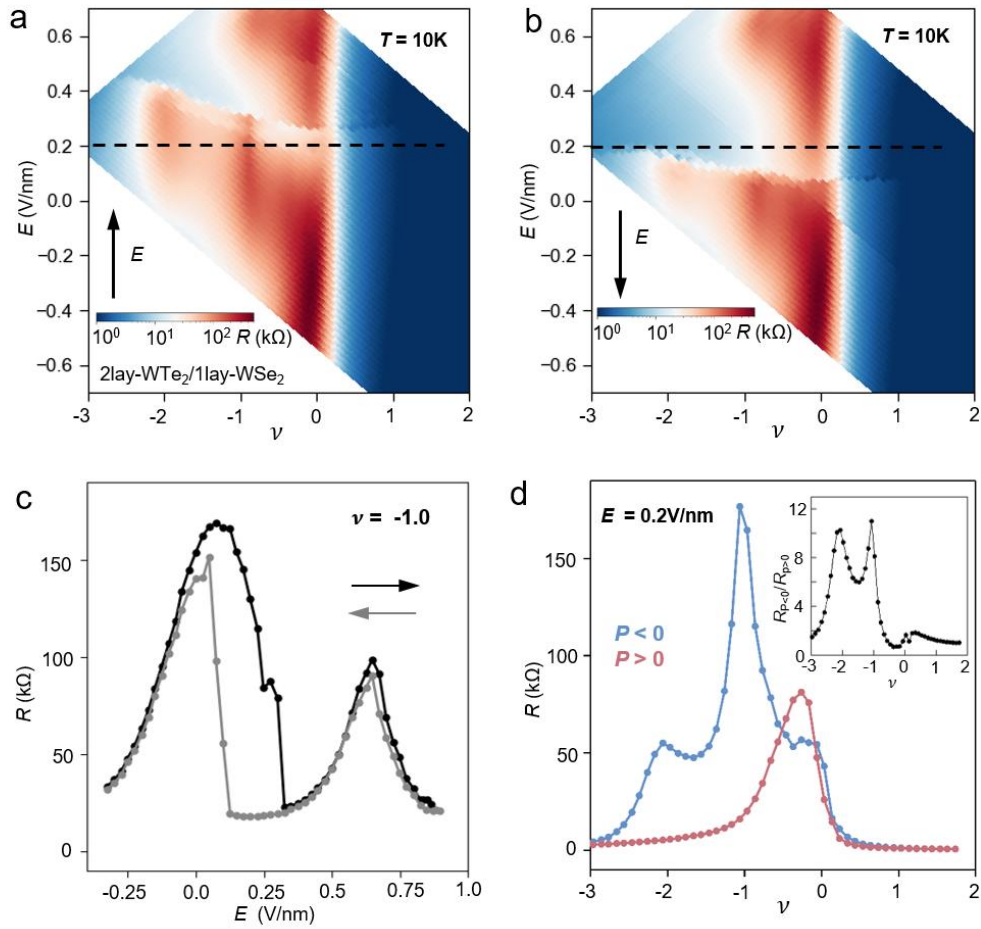

**Figure 2 | Switching moiré potential by ferroelectricity. a,b**, Electric field and filling factor dependence for the longitudinal resistance of an angle-aligned bilayer $T_d$-$WTe_2$/monolayer H-$WSe_2$ heterostructure at 10 K. The black arrows label the forward (**a**) and backward (**b**) electric field scan directions. Hysteretic electric field dependence corresponding to ferroelectric switching is observed. **c**, Electric field dependence of the longitudinal resistance at $\nu = -1$. The arrows mark the field scan directions. **d**, Filling factor dependent longitudinal resistance extracted from **a** (blue) and **b** (red) at $E = 0.2$ V/nm (the black dashed lines in **a,b**). They denote the two spontaneous polarization states $P > 0$ and $P < 0$. Moiré insulating states are observed only for $P < 0$. Inset: filling factor dependent resistance on/off ratio $R_{P<0}/R_{P>0}$.

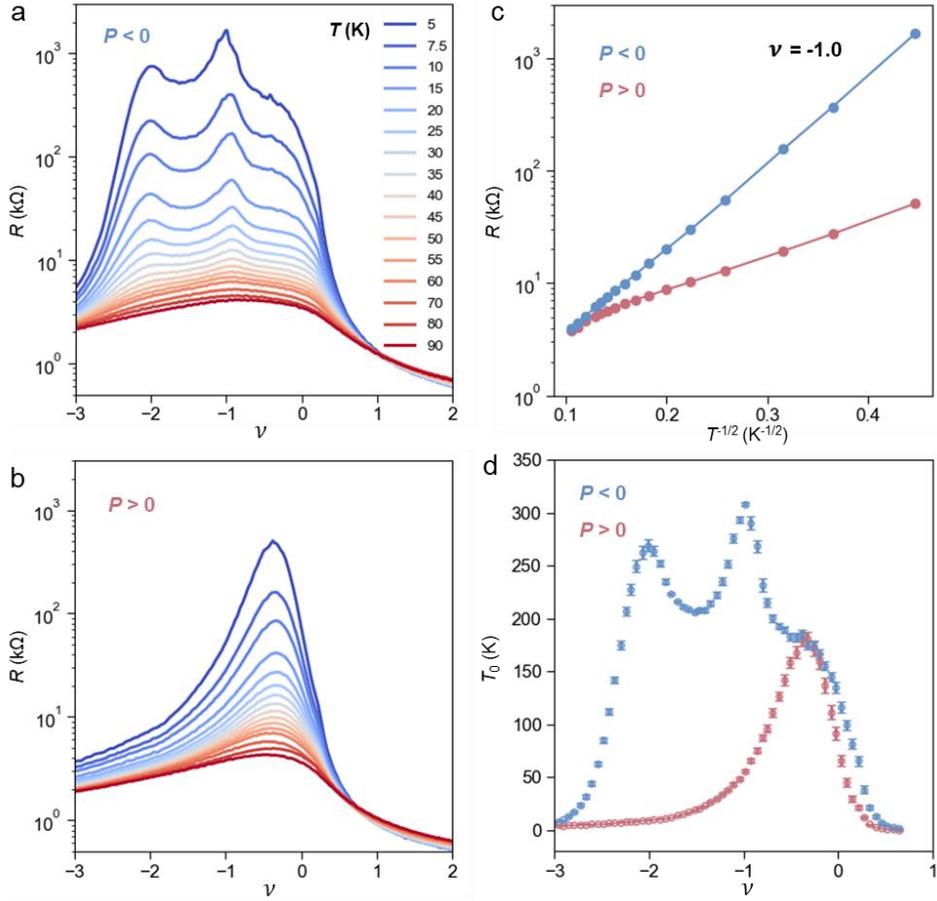

**Figure 3 | Temperature dependent electrical transport. a,b**, Filling factor dependent longitudinal resistance at varying temperatures from 5 K to 90 K for $P < 0$ (**a**) and $P > 0$ (**b**). Insulating-like behavior is observed for the entire hole-doping region in both cases. **c**, Resistance (in log scale) versus $T^{-1/2}$ at $\nu = -1$ for both $P < 0$ and $P > 0$. The linear dependence demonstrates the Efros-Shklovskii variable-range hopping. **d**, Filling factor dependence of the extracted temperature $T_0$ for both $P < 0$ and $P > 0$. Significant enhancements at the moiré insulating states are observed, showing the shortened localization lengths when the moiré potential is turned on.

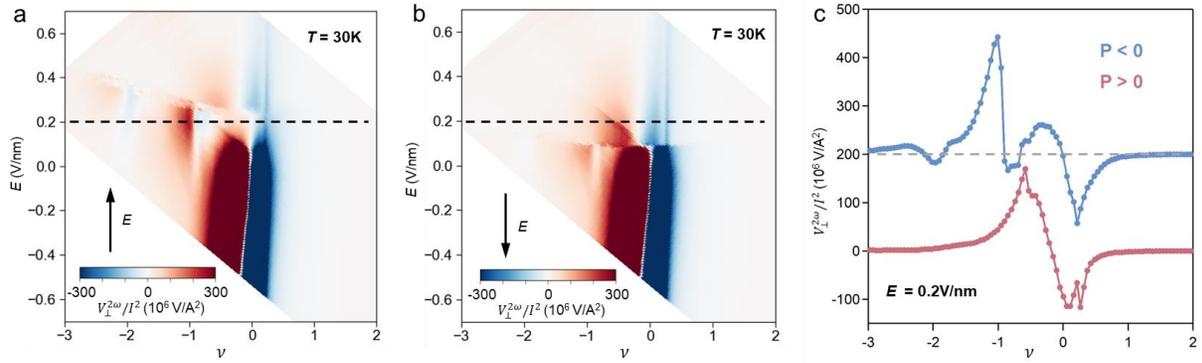

**Figure 4 | Switching the nonlinear anomalous Hall transport by ferroelectricity. a, b**, Electric field and filling factor dependent nonlinear anomalous Hall response $V_\perp^{2\omega}/I^2$ for the forward (**a**) and backward (**b**) electric field scan directions (black arrows). Hysteretic electric field dependence corresponding to ferroelectric switching is observed. **c**, Filling factor dependent $V_\perp^{2\omega}/I^2$ for $P < 0$ and $P > 0$ extracted from **a** (blue) and **b** (red), respectively, at $E = 0.2$ V/nm (the black dashed lines in **a,b**). The two curves are vertically displaced for clarity. The horizontal dashed line marks $V_\perp^{2\omega}/I^2 = 0$ for $P < 0$. Nonlinear anomalous Hall resets are observed at the moiré insulating states at $\nu = -1, -2$ only for $P < 0$.

**Extended Data Figures**

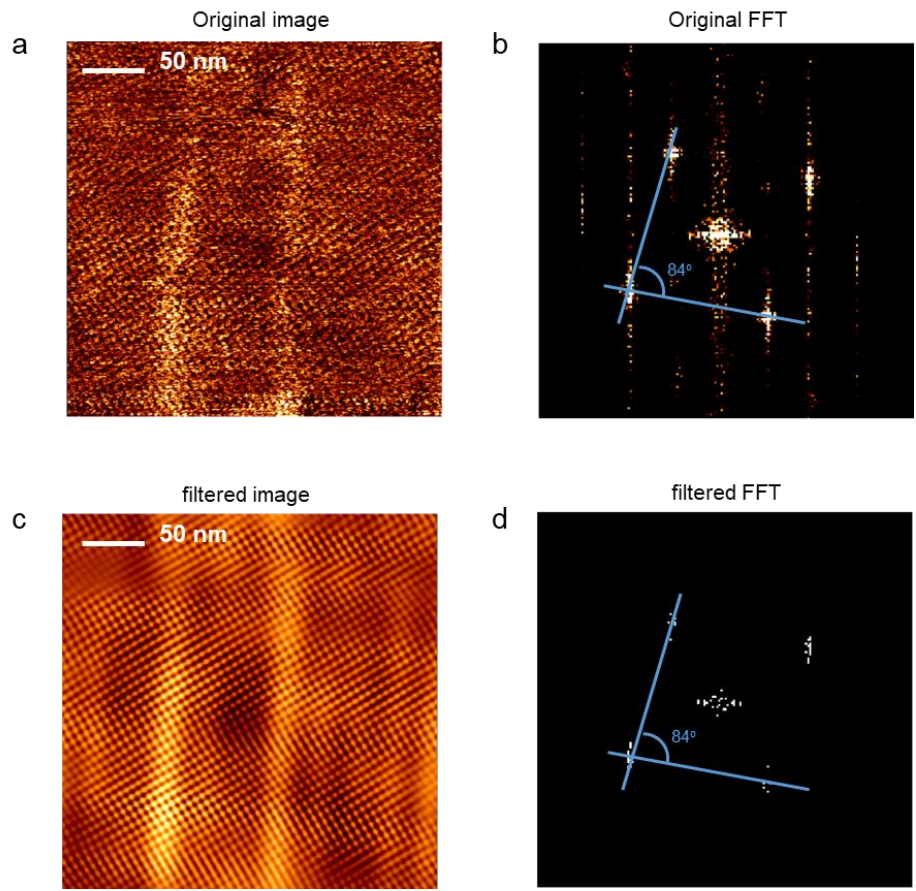

**Extended Data Figure 1 | PFM images and analysis. a,** Raw PFM image of an angle-aligned monolayer $T_d$-WTe$_2$/monolayer H-WSe$_2$ heterostructure. **b,** Fourier transform of the raw image. Fourier peaks corresponding to a centered-rectangular lattice can be clearly observed. **c, d,** The Fourier-filtered PFM image (**c**) and the corresponding Fourier transform (**d**). A high-pass filter is applied with a threshold of 72 % of the maximum intensity.

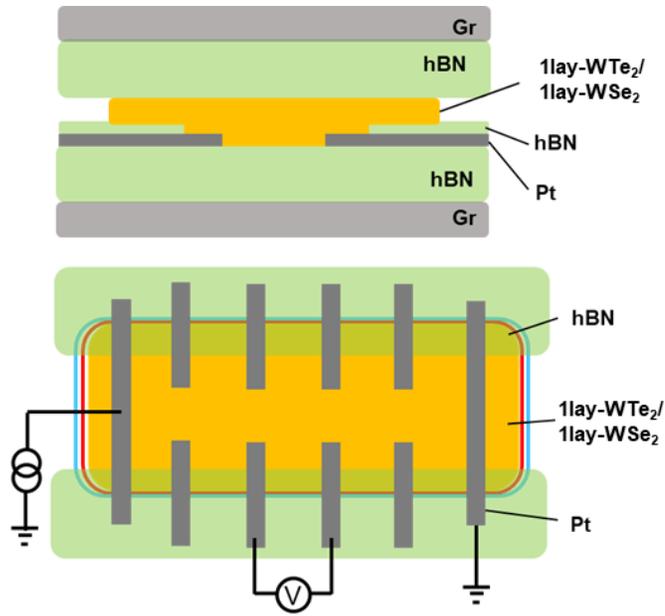

**Extended Data Figure 2 | Schematics of an angle-aligned monolayer $T_d$-WTe$_2$/monolayer H-WSe$_2$ Corbino device.** Top: cross-section of the device. A thin hBN spacer is inserted in between the WTe$_2$ and the platinum (Pt) electrodes to avoid direct edge contacts. Gr stands for few-layer graphite gate electrode. Bottom: top view of the device. The blue and red lines denote the helical edge states of a quantum spin Hall insulator. The measurement configuration is also shown.

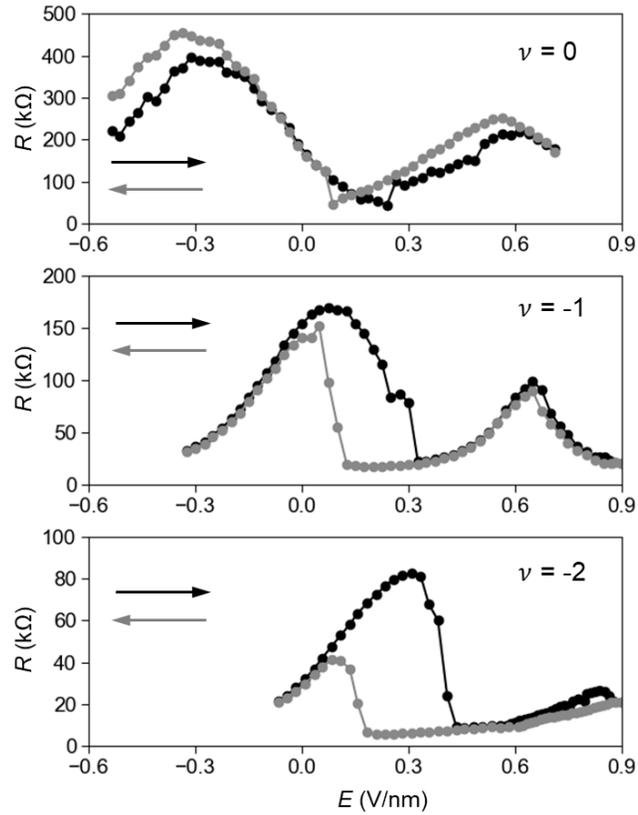

**Extended Data Figure 3 | Electric field dependent resistance at selected filling factors of $\nu = 0, -1, -2$.** Arrows denote the electric field scan directions. The measurement temperature is 10 K. A clear hysteresis corresponding to ferroelectric switching is observed. In addition to the ferroelectric switching, non-monotonic electric field dependence is also observed (see discussions in Methods).

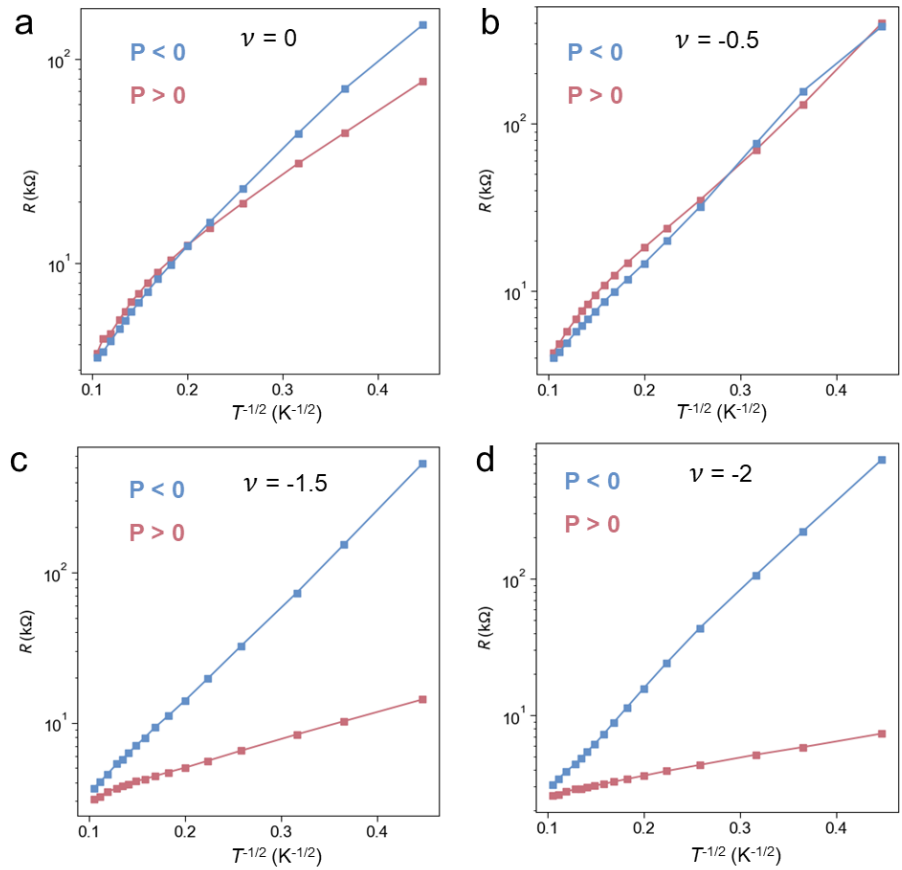

**Extended Data Figure 4 | Variable-range hopping transport at different filling factors. a-d,** Resistance (in log scale) versus $T^{-1/2}$ for both $P < 0$ and $P > 0$ at $\nu = 0$ (**a**), -0.5 (**b**), -1.5 (**c**) and -2 (**d**). The linear dependence demonstrates the Efros-Shklovskii variable-range hopping.

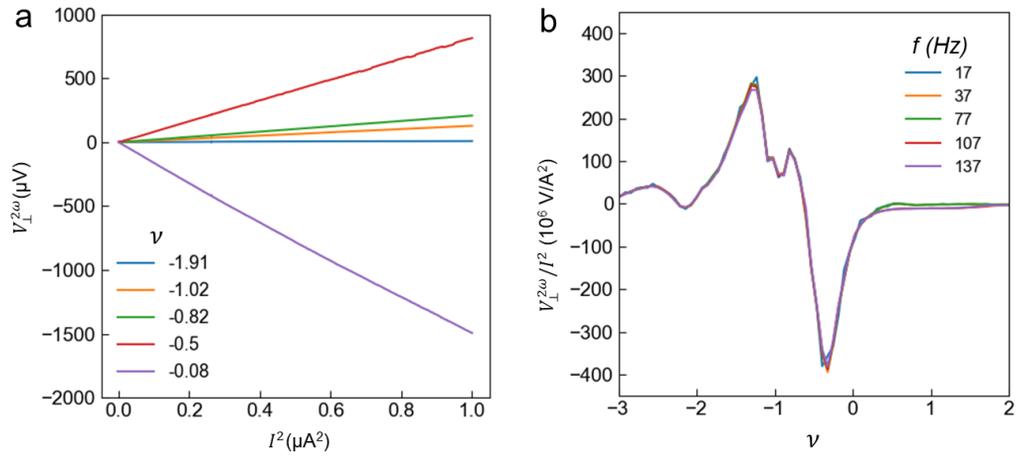

**Extended Data Figure 5 | Nonlinear anomalous Hall response. a,** Linear dependence of the second-harmonic Hall voltage on the bias current squared at varying filling factors. The current modulation frequency is 17 Hz. The measurement temperature is 25 K. **b,** Filling factor dependence of the nonlinear anomalous Hall response $V_\perp^{2\omega}/I^2$ at zero bottom gate voltage and at varying excitation frequencies of 17, 37, 77, 107, and 137 Hz. The response is independent of the excitation frequency.

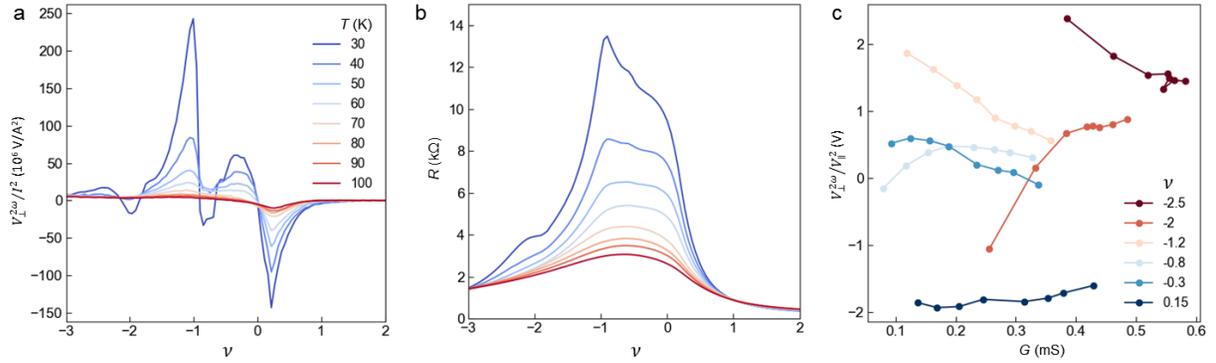

**Extended Data Figure 6 | Temperature dependent nonlinear anomalous Hall effect at $E = 0.2$ V/nm and $P < 0$. a,b,** Filling factor dependence of $\frac{V_\perp^{2\omega}}{I^2}$ (**a**) and the longitudinal resistance $R$ (**b**) at varying temperatures from 30 K to 100 K. **c**, Extracted $V_\perp^{2\omega}/V_\parallel^2$ as a function of the sample conductance $G$ at selected moiré filling factors from –2.5, to 0.15. Unlike the extrinsic NAHE in the coherent metallic transport regime, in which $V_\perp^{2\omega}/V_\parallel^2 \propto G^2$ is expected and has been observed [25], complicated dependence of $V_\perp^{2\omega}/V_\parallel^2$ on $G$ is observed for the variable-range hopping transport regime here. Future studies are required to better understand the NAHE in the hopping transport regime.

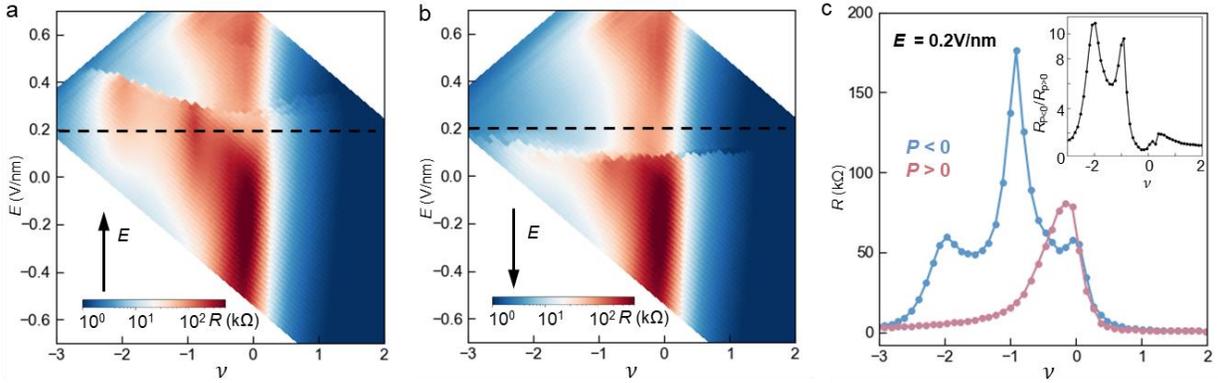

**Extended Data Figure 7 | Transport studies along the WTe$_2$ crystal b-axis. a,b,** Electric field and filling factor dependence for the longitudinal resistance of angle-aligned bilayer T$_d$-WTe$_2$/monolayer H-WSe$_2$ heterostructure at 10 K. The current is biased along the crystal b-axis of WTe$_2$. The black arrows label the forward (**a**) and backward (**b**) electric field scan directions. Hysteretic electric field dependence corresponding to ferroelectric switching is observed. **c,** Filling factor dependent longitudinal resistance extracted from **a** (blue) and **b** (red) at $E = 0.2$ V/nm (the black dashed lines in **a,b**). They denote the two spontaneous polarization states $P > 0$ and $P < 0$. Moiré insulating states are observed only for $P < 0$. Inset: filling factor dependent resistance on/off ratio $R_{P<0}/R_{P>0}$.